\documentclass{article} 
\usepackage{iclr2020_conference,times}

\usepackage{amsmath,amsfonts,bm}

\def\eqref#1{equation~\ref{#1}}

\def\1{\bm{1}}

\DeclareMathAlphabet{\mathsfit}{\encodingdefault}{\sfdefault}{m}{sl}
\SetMathAlphabet{\mathsfit}{bold}{\encodingdefault}{\sfdefault}{bx}{n}

\usepackage{hyperref}
\usepackage{url}

\usepackage{tikz}
\usetikzlibrary{arrows,positioning} 
\tikzset{
    >=stealth',
    punkr/.style={
           rectangle,
           rounded corners,
           draw=black, very thick,
           text centered},
    punkt/.style={
           circle,
           draw=black, very thick,
           text width=4em,
           text centered},
    pil/.style={
           ->,
           very thick,
           shorten <=4pt,
           shorten >=4pt,}
}

\title{An Overview and Case Study of the Clinical AI Model Development Life Cycle for \\Healthcare Systems}

\author{Charles Lu,\; Julia Strout,\; Romane Gauriau,\; Brad Wright, \\
\textbf{Fabiola Bezerra De Carvalho Marcruz,\; Varun Buch\; \&\; Katherine Andriole}  \\
Massachusetts General Hospital \\
Brigham and Women's Hospital \\
Harvard Medical School \\
\texttt{clu@mgh.harvard.edu} \\
}

\iclrfinalcopy 
\begin{document}

\maketitle

\begin{abstract}
Healthcare is one of the most promising areas for machine learning models to make a positive impact. However, successful adoption of AI-based systems in healthcare depends on engaging and educating stakeholders from diverse backgrounds about the development process of AI models. We present a broadly accessible overview of the development life cycle of clinical AI models that is general enough to be adapted to most machine learning projects, and then give an in-depth case study of the development process of a deep learning based system to detect aortic aneurysms in Computed Tomography (CT) exams. We hope other healthcare institutions and clinical practitioners find the insights we share about the development process useful in informing their own model development efforts and to increase the likelihood of successful deployment and integration of AI in healthcare.

\end{abstract}

\section{Introduction}
The field of machine learning (ML) has the potential to fundamentally improve healthcare systems by capitalizing on the advances of deep learning in computer vision, natural language processing, and speech recognition, facilitated by the increasing accumulation of medical data and the widening availability of computing resources \citep{Krizhevsky, Devlin, Chorowski}. There continues to be an increasing amount of research in using models to predict diseases, detect biomarkers, improve patient triage, decrease diagnosis time, facilitate novel drug discovery, and optimize hospital operations \citep{Fauw, Esteva, Dana, Putin, Annarumma}.
However, in spite of all this promising research, a divide remains between model development and successfully translating results into actual clinical application. Machine learning in healthcare differs from many other domains, in that labeled data is extremely expensive to obtain as it must be labeled by clinicians and deployment to legacy hospital systems requires tight vendor integration in a highly regulated environment. Clinicians should critically and carefully assess AI applications, as many previous attempts (e.g. expert systems and IBM Watson), have had difficulty meeting initial expectations in medicine \citep{Heathfield, Strickland}.

For clinical AI to be adopted successfully, working processes need to be formalized to help mitigate risk while developing AI-related projects; however, existing project methodologies from software engineering are difficult to adapt to clinical ML projects, which are more similar to applied research projects, where progress and final deliverables are difficult to estimate \emph{a priori}. Model development is an iterative process which depends heavily on prototypes, experimentations, and constant interaction with clinical expertise to guide and interpret results. To our knowledge, there is a lack of existing work that focuses on a task-agnostic ML model development framework for clinical applications in healthcare systems.

By detailing our clinical AI development process and extrapolating common differentiators of successful projects, we hope other organizations can incorporate useful aspects to bridge the gap between machine learning and clinical utility. In the following sections, we first present a detailed overview of the development cycle for clinical AI projects aimed at a broad audience and then give a model development case study of a project to detect aortic aneurysms. 

\section{The clinical AI development cycle}

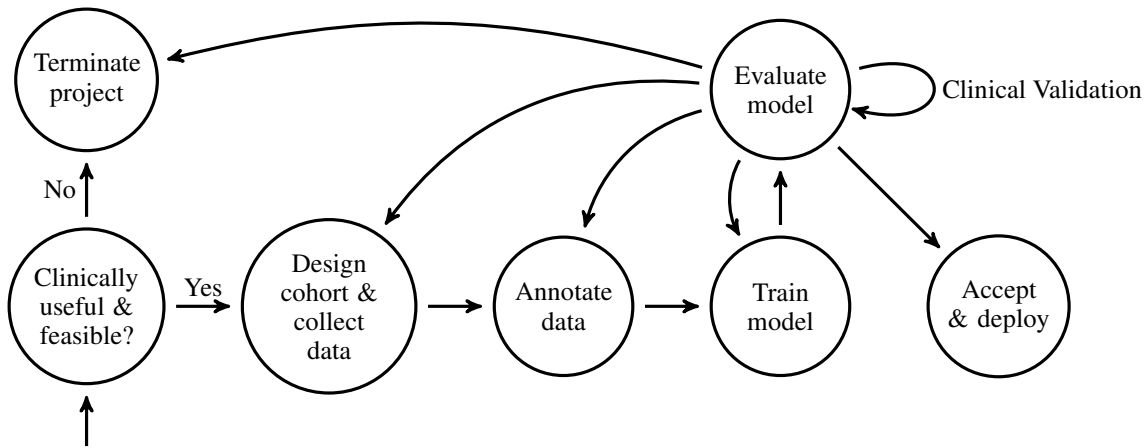
\begin{figure}[h]
\begin{center}
\begin{tikzpicture}[node distance=1cm, auto,]
	\node[punkt] (a) {Clinically useful \& feasible?};
	\node[punkt, right=of a] (b) {Design cohort \& collect data};
	\node[punkt, right=of b] (c) {Annotate data};
	\node[punkt, right=of c] (d) {Train model};
	\node[punkt, above=of d] (e) {Evaluate model};
	\node[punkt, right=of d] (f) {Accept \& deploy};
	\node[punkt, above=of a] (g) {Terminate project};
	\path
	(a) edge[pil] node {Yes} (b)
	(b) edge[pil] node {} (c)
	(c) edge[pil] node {} (d)
	(d) edge[pil] node {} (e)
	(e) edge[pil] node {} (f)
	(e) edge[pil, loop right] node {Clinical Validation} (e)
	(e) edge[pil, bend right=30] node {} (d)
	(e) edge[pil, bend right=30] node {} (c)
	(e) edge[pil, bend right=30] node {} (b)
	(e) edge[pil, bend right=15] node {} (g) 
	(a) edge[pil] node {No} (g);
	\draw[pil, <-, very thick, shorten <=4pt, shorten >=4pt] (a) --++(270:2cm);
\end{tikzpicture} 
\end{center}
\caption{The clinical AI model development cycle.}
\end{figure}

\subsection{Feasibility and impact assessment of project}
Before undertaking any development efforts, a proposed project should be evaluated for feasibility, scope, and clinical impact. Representatives from all stakeholders including clinical, technical, and commercial teams should be present to establish reasonable goals and timelines for the project. Potential projects should have a clear, well-defined clinical use case for the model's application. In our experience, most project proposals do not pass the feasibility and impact assessment stage. A common reason why a project is not undertaken is due to insufficient data resources, either lacking the quantity or quality to train a performant model. Other projects may receive a low priority if the clinical impact is too narrow or limited in scope, such as a model to predict rare or non-critical conditions.

If a project is deemed feasible and impactful, important design decisions must be formalized that will impact the rest of the development cycle. Considerations at this stage can include defining valid inputs to the model, choosing what labels to collect or annotations to perform, strategies for clinical evaluation, aligning expectations on timelines, and defining the acceptance criteria.

\subsection{Data acquisition, cohort selection, and data cleaning}
Once the feasibility and impact of a project has been assessed, the first step for most projects is the collection and aggregation of data resources. Within a healthcare system, several teams must coordinate and take measures to ensure that Protected Health information (PHI) is safeguarded and privacy is maintained. This may require de-identification of medical data and obtaining the requisite permissions and approvals from human research committees and Institutional Review Boards (IRB).

After the bulk data is acquired, a cohort must be selected while considering multiple factors such as patient demographics, exclusion criteria, and even characteristics of the data acquisition process itself. Especially in medical data, a long tail of outlying edge cases may need to be excluded, and qualifications may need to be placed as to what is considered valid for the model to accept as input. Biases in the data acquisition process itself can also create issues during model development and evaluation. For example, in one of our projects to detect stroke in CT, we realized that an imbalance in our dataset acquired through imaging scanners of two different manufactures caused the model to make negative predictions 50\%-50\% but output positive predictions 95\%-5\% skewed towards the predominant manufacture! This examples highlights the need to balance multiple factors of the training dataset to learn a model that generalizes properly across varying data sources. Determining more subtle parameters of the data will usually require the domain expertise of a clinician. 

Once data is acquired and a cohort is created, additional processing, cleaning, normalization, and inspection of the data is done to standardize the format before annotating the data and training the model. These tasks might comprise of imputing missing data, correcting erroneous data, resolving inconsistencies in data, filtering extraneous data, excluding outliers in the data, and checking the data for quality assurance and correctness.

\subsection{Data annotation and labeling}
For any supervised learning task, the dataset must have accompanying labels to train a model and to evaluate performance. Determining the type and granularity of labeling depends on the clinical use case and model learning task. Sometimes, it is possible to leverage passive annotation methods to extract labels from existing structured data such as parsing ICD-10 medical codes or using natural language processing to retrieve relevant keywords in the patient reports. Otherwise manual annotations must be collected. 

One unique aspect of developing clinical AI models is that annotation cannot typically be done without considerable domain expertise and training, such as identifying Ventricular Tachycardia in Electrocardiograms (ECG) data or localizing Large Vessel Occlusions in CT Angiograms. Examples of other annotation tasks can include identifying abnormal conditions or events, measuring the volume of anatomical structures, and placing or drawing markers (points, bounding boxes, polygons) around regions of interest. While being considerate of the clinician's time, medical image annotation is often a resource-intensive bottleneck in the model development cycle.

We recommend performing a trial of annotating a small subset of the dataset to check for consistency and correctness among annotators before annotating the entire dataset. In our experience, capturing excellent annotations will prevent many issues downstream in the development process as poor annotations lead to a subpar model and difficulties in evaluating model performance on the ground truth. If possible, ambiguous or difficult cases should have multiple annotators to ensure consistent ground truth labels. Trials to measure inter-annotator variability could also be conducted as clinicians may interpret the same case differently due to differences in training, expertise, and other factors. 

\subsection{Model exploration, development, and training}
Once annotations are complete, the labeled dataset should be split into training and validation sets for development of the model and a test set for final evaluation of the model. A baseline model will first be developed to gauge the difficulty of the problem and give some indication of the performance improvement necessary to reach the acceptance criteria. When a baseline has been established, efforts can be made to improve the model's performance through changing the model architecture, further data processing, feature engineering, data augmentation strategies, model ensembling techniques, and hyper-parameter tuning. In practice, we found that engineering effort is usually well spent on developing a robust data processing pipeline and focusing on techniques to increase signal to the model, for example registering the skull to a standard atlas and skull-stripping before feeding into an inter-cranial stroke classification model.

Other model design decisions are driven by specifications of the clinical use case. In some cases, even if a model does not attain high performance on some metrics, it may still provide clinical utility (e.g. clinical workflow improvements), while models with higher performance may not actually be clinically useful if they are not robust to variations in the data distribution or require extremely long inference times. Also, some types of errors are more egregious than others for the clinical application, and the use case should inform model development priorities, such as setting the ROC operating point for the appropriate sensitivity versus specificity of a classifier.

\subsection{Model testing and clinical evaluation}
After a model has been trained and meets the performance metrics on the validation set, it can be evaluated on the holdout test set as a measure of its ability to generalize on the unseen data (i.e. true data distribution). Crucially, once a test set has been evaluated, an additional test set must be collected for further development to avoid model bias. If a large discrepancy exists between the performance on the validation set and the test set, it may be due to \emph{data leakage} (when information about the validation set is passed to the model, which results in an overestimation of the model's generalizability) or the test set having a disparate distribution from the validation set, which can occur with small samples sizes or when the dataset was not stratified properly. Care must be taken to ensure that multiple images or data points for the same patient do not cross between training, validation, and test sets. In cases where the dataset is extremely limited or imbalanced, other data partitioning schemes such as \emph{k-fold cross validation} or stratified sampling can be applied.

Besides measuring performance metrics, the results of the model's predictions should also be clinically evaluated for failure modes and clinical mimics. An analysis of the types and frequencies of prediction errors may reveal interesting insights about the dataset or labels, such as incorrect ground truth, corrupted data, or hidden biases in the data cohort, that can used to improve the model or adjust the clinical use case. Inter-clinician experiments can also be conducted to measure agreement between the model and the clinician as well as agreement between clinicians.

\subsection{Acceptance, termination, or iteration}
If the model does not meet the performance requirements of the clinical use case and development cannot continue, the project may terminate at this stage, in which case a post mortem of the life cycle can be conducted. If the model meets performance criteria and is accepted into the deployment cycle, next steps can include additional clinical evaluation such as external off-site validation, conducting clinical trials to submit for FDA approval, or gathering application feedback in a clinical environment. Often, projects continue iterating into previous stages multiple times during the course of a project.

\section{Aortic aneurysm model case study}
Aortic aneurysms are abnormal enlargements in the wall of the aorta, the major blood vessel that connects the heart to the rest of the body, which can fatally rupture with a 90\% mortality rate if left untreated. Because most aneurysms are asymptomatic until rupturing and identified incidentally only 65\% of time, there is clinical utility in developing a model to automatically detect aortic aneurysms on routine chest and abdominal CT examinations \cite{Claridge}. The clinical definition for aortic aneurysms is based on the measurement of the largest diameter of the aorta, so the model has to perform several subtasks (segmentation, measurement, classification) to produce the needed outputs for clinical application. Our project proceeded through three iterations of the development life cycle detailed below. 

\begin{figure}[h]
\begin{center}
\begin{tikzpicture}[node distance=1cm, auto,]
	\node[punkr] (a) {Assess project};
	\node[punkr, fill=red!60!gray, right=of a] (b) {Collect data};
	\node[punkr, fill=red!60!gray, right=of b] (c) {Annotate};
	\node[punkr, fill=red!60!gray, right=of c] (d) {Train model};
	\node[punkr, fill=red!60!gray, right=of d] (e) {Evaluate};
	\node[punkr, fill=orange!60!gray, below=of e] (f) {Collect more data};
	\node[punkr, fill=orange!70!gray, left=of f] (g) {Annotate data};
	\node[punkr, fill=orange!70!gray, left=of g] (h) {Train model};
	\node[punkr, fill=orange!70!gray, left=of h] (i) {Evaluate};
	\node[punkr, fill=yellow!70!gray, below=of i] (j) {Annotate data};
	\node[punkr, fill=yellow!70!gray, right=of j] (k) {Train model};
	\node[punkr, fill=yellow!70!gray, right=of k] (l) {Evaluate};
	\node[punkr, fill=green!55!gray, right=of l] (m) {Accept model};
	
	\path
	(a) edge[pil] node {} (b)
	(b) edge[pil] node {} (c)
	(c) edge[pil] node {} (d)
	(d) edge[pil] node {} (e)
	(e) edge[pil] node {} (f)
	(f) edge[pil] node {} (g)
	(g) edge[pil] node {} (h)
	(h) edge[pil] node {} (i)
	(i) edge[pil] node {} (j)
	(j) edge[pil] node {} (k)
	(k) edge[pil] node {} (l)
	(l) edge[pil] node {} (m);

\end{tikzpicture}
\end{center}
\caption{The aortic aneurysm model life cycle. Colors correspond to development iteration.}
\end{figure}
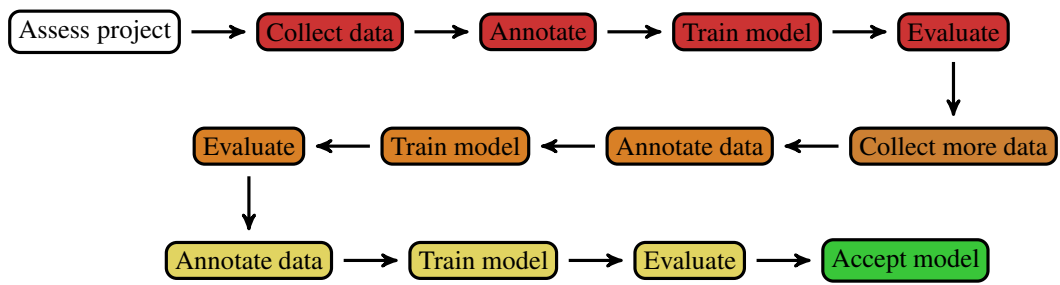

\subsection{First iteration}
After the clinical use case was defined and project investigators obtained IRB approval, a data acquisition pull was requested for 1000 contrast and non-contrast CT examinations of the chest, abdomen, and pelvis that matched keywords relating to aortic aneurysms in the radiological patient reports. After the data was pulled, the raw patient-level data had to be filtered to contain only the appropriate examination-level images for each patient as an individual patient often has multiple examinations. Exams were further filtered by acquisition parameters of the imaging scanner, such as slice thickness and number of slices in an exam. This selection process was only partially automated as examination metadata is not always reliable and often missing or incorrect. Manual inspection for image quality and confirmation of the presence of aneurysm was also performed by radiologists. Overall, this data curation process took several months before annotations could start.

For purposes of visualization and aneurysm measurement, annotations consisted of volumetric 3D segmentation masks of the entire aorta, including the any present aneurysms or pseudo-aneurysms. Other labels, including the location and diameter measurement of the aneurysm, were extracted from the radiological report data. Manually segmenting the aorta is a time consuming but not a very clinically challenging task, so an external party of medically trained annotators were used, which required de-identification of all PHI in the examination data.

The primary model architecture was a 3D U-Net-like model, similar to \cite{Ronneberger}, to output a segmentation prediction of the abdominal aorta from abdominal-pelvic CT exams. This predicted segmentation volume was post-processed and an ellipse fitting algorithm extracted the maximal diameter perpendicular to the centerline axis of the aorta. See our paper \cite{Lu} for more information. The abdominal model achieved good performance on the test set with Dice score of 0.9, sensitivity of 91\%, and specificity of 95\%. This development iteration required 9 months.

\subsection{Second iteration}
Secondary development of a model to detect thoracic aortic aneurysms using a similar approach proved to be much more difficult as the thoracic aorta curves in a cane-like shape through the chest cavity while the abdominal aorta is predominately straight. Upon review of the segmentation annotations of the thoracic aorta, we uncovered quality issues, including missing sections of the segmented aorta, over-annotation of the aortic root, and under-annotation of the aortic arch. To address these annotation quality issues, we collected a new batch of data to annotate with additional training of the annotators to ensure a better standard level of quality. This added 3 additional months to the project timeline.

After the new annotations were complete, an expert clinician at our institution checked the annotations for quality and made minor corrections. Additional engineering effort was devoted to new morphological processing and angle correction algorithms to extract the correct diameter measurements from the thoracic aorta segmentations. These diameter measurements would be used to train an ensemble classifier for a final prediction. The combination of higher quality segmentation labels, improved post-processing transformations, and additional classification model led to evaluation performance that exceeded the acceptance criteria with a Dice score of 0.93, sensitivity of 96\%, and specificity of 95\%.

\subsection{Third iteration}
A final iteration cycle was needed to develop a model to delineate the boundary between thorax and abdomen during inference time to route the appropriate CT exam to the correct segmentation model. Annotations were reasonably cheap as only a slice-level label denoting the boundary was needed for each exam. A model based off the ResNet architecture from \cite{He} was trained to regress slice-level and output a scalar value predicting the distance from the boundary. Then, portions of the image containing the thoracic aorta would be sent the thoracic model while sections of the abdominal aorta would be sent to the abdominal model for prediction.

After clinical evaluation and alpha testing of 1000 additional retrospective cases, all models and their dependencies were packaged to be integrated into existing clinical workflow software. With integration complete, a beta test in a clinical environment was scheduled before wider deployment. 

\bibliography{iclr2020_conference}
\bibliographystyle{iclr2020_conference}

\appendix

\section{The deployment cycle}
While we focused mainly on the development cycle, we briefly discuss the deployment cycle. Once a model has been through evaluation and met the acceptance criteria, the model must be packaged and ingested in an inference system to be integrated into the clinician workflow. This process can be quite involved and domain specific depending on the inference platform and integration requirements for existing point-of-care software and database systems, often requiring coordination between several external vendors and regulatory committees. 

For some of our projects, the next phase after model development is \emph{shadow inference}, where the model runs inference on live data streams but does not impact clinical decision making. Shadow inference is performed to test system integration and gather feedback on the clinical application. Ongoing efforts are made to interface with clinicians to develop standards for deployment of clinical AI applications. Other important considerations of model deployment include continual monitoring for different failure modes and distribution shifts in non-stationary data. As clinical decisions are influence by models predictions, models will need to be continuously evaluated, recalibrated, and retrained on new data as the population changes over time. Other interesting work around model deployment includes federated learning, clinician-machine interfaces, and active learning techniques.

\end{document}